\documentclass[journal]{IEEEtran}

\usepackage[english]{babel}
\usepackage{blindtext}
\usepackage{cite}

\usepackage[inline]{enumitem}
\usepackage{lipsum}
\usepackage[table]{xcolor}

\usepackage{xurl}

\usepackage{xspace}
\usepackage{etoolbox}
\usepackage{xstring}
\usepackage{array}
\usepackage[flushleft]{threeparttable}
\usepackage{listings}

\usepackage{amssymb}
\usepackage{pifont}
\usepackage{multirow}
\usepackage{listings}
\usepackage{parcolumns}
\usepackage{caption}
\usepackage{subcaption}
\usepackage{textcomp,amsmath}
\usepackage{adjustbox}
\usepackage{booktabs}

\DeclareCaptionFont{white}{\color{white}}
\DeclareCaptionFormat{listing}{\colorbox{gray}{\parbox{0.47\textwidth}{\footnotesize#1#2#3}}}
\usepackage{tikz}
\usepackage{graphicx}
\usepackage{xspace}

\DeclareListParser{\doslashlist}{/}
\newcounter{ndnNameComponentCounter}%
\newcommand{\name}[1]{{%
		\setcounter{ndnNameComponentCounter}{0}%
		\renewcommand{\do}[1]{{%
				\ifnumgreater{\value{ndnNameComponentCounter}}{0}{\allowbreak/}{}%
				\ifnumodd{\value{ndnNameComponentCounter}}{}{}%
				\detokenize{##1}}%
			\stepcounter{ndnNameComponentCounter}}%
		``{\fontfamily{cmtt}\small\selectfont\IfBeginWith{#1}{/}{/}{}\doslashlist{#1}}''%
}}

\newcommand{\ndnname}[1]{{\footnotesize {#1}}\xspace}%

\usepackage{soul}
\usepackage{ifthen}

\newboolean{showcomments}
\setboolean{showcomments}{true}
\ifthenelse{\boolean{showcomments}}
{ \newcommand{\mynote}[3]{
		\protect\fbox{\bfseries\sffamily\tiny#1}
		{\small$\blacktriangleright$\textsf{\emph{\color{#3}{#2}}}$\blacktriangleleft$}}}
{ \newcommand{\mynote}[3]{}}

\newcommand{\outline}[1]{}

\def\first{({i})\xspace}
\def\second{({ii})\xspace}
\def\third{({iii})\xspace}

\newcommand{\sysname}{Sovereign\xspace}

\newcommand{\cmark}{\ding{51}}%
\newcommand{\xmark}{\ding{55}}%

\newcommand{\eg}{{\it e.g.,}\xspace}%
\newcommand{\ie}{{\it i.e.,}\xspace}%
\newcommand{\etal}{{\it et.~al}\xspace}%
\newcommand{\mypara}[1]{\smallskip\noindent{\bf {#1}:}~}%

\newif\ifmark%
\newif\ifhidenote%
\marktrue%
\hidenotetrue%

\newif\ifspace%
\spacetrue%

\usepackage[normalem]{ulem}
\ifmark
\newcommand{\del}[1]{\sout{#1}}%
\ifhidenote
\newcommand{\note}[1]{}%
\else
\newcommand{\note}[1]{{\sffamily\itshape\bfseries\uline{#1}}}%
\fi
\else
\newcommand{\del}[1]{}%
\newcommand{\note}[1]{}%
\fi

\definecolor{__red}{rgb}{0.8,0.1,0.1}
\definecolor{__green}{rgb}{0.1,0.6,0.1}

\definecolor{verylightgray}{gray}{0.8}
\newcommand{\yestick}{{\color{__green}{\cmark}}}
\newcommand{\notick}{{\color{__red}{\xmark}}}

\begin{document}
%
\title{\sysname: Self-contained Smart Home with Data-centric Network and Security}
%
%
%

\author{Zhiyi Zhang,~\IEEEmembership{Member,~IEEE,}
        Tianyuan Yu,
        Xinyu Ma,
        Yu Guan,
        Philipp Moll,
        and~Lixia Zhang,~\IEEEmembership{Fellow,~IEEE}
\thanks{Zhiyi Zhang, Tianyuan Yu, Xinyu Ma, Philipp Moll, and Lixia Zhang are with the Computer Science Department, UCLA. Email addresses: zhiyi, tianyuan, xinyu.ma, phmoll, lixia@cs.ucla.edu}
\thanks{Yu Guan is with Peking University. Email address: shanxigy@pku.edu.cn}}

\maketitle

\begin{abstract}

Recent years have witnessed the rapid deployment of smart homes; most of them are controlled by remote servers in the cloud.
Such designs raise security and privacy concerns for end users.
In this paper, we describe the design of \sysname, a home IoT system framework that provides end users complete control of their home IoT systems.
\sysname lets home IoT devices and applications communicate via application-named data and secures data directly.
This enables direct, secure, one-to-one and one-to-many device-to-device communication over wireless broadcast media.
\sysname utilizes semantic names to construct usable security solutions.
We implement \sysname as a publish-subscribe-based development platform together with a prototype home IoT controller.
Our preliminary evaluation shows that \sysname provides a systematic, easy-to-use solution to user-controlled, self-contained smart homes running on existing IoT hardware without imposing noticeable overhead.

%
\end{abstract}


%
\IEEEpeerreviewmaketitle

\section{Introduction}

Technology advances lead to improvements in home living.
For example, appliances such as dishwashers and refrigerators, which are directly controlled by home users, improve the quality of life at home.
Over the last decade, Internet-of-Things (IoT) has come of age~\cite{smart-home-statistics}, and networked smart devices at home lead to home reinvention.
However, different from conventional home appliances, most of today's popular smart home deployments~\cite{smartthings, googlenest, awsiot, homekit, microsoftiot, alibabaiot} are controlled by remote servers in the cloud.

A fundamental concern with today's smart home deployment practice is the exposure of users' daily home life operations to cloud service providers~\cite{tweneboah2017cyber, lin2016iot, ur2013current, choe2011living, shang2017breaking}.
In today's smart homes, devices are controlled by the cloud.
To control one's home, a user contacts the cloud backend, which authenticates the user, and then authorizes him/her to issue commands from the cloud to his/her home appliances.
This effectively presents a user only as an \emph{authorized client} of her own home rather than the owner.

To this end, many works have been devoted to higher user privacy from the cloud service providers and stronger user control of home devices and data~\cite{jedi, trust-but-verify, jia2017contexlot, tian2017smartauth, demetriou2017hanguard, celik2019iotguard, yahyazadeh2019expat, simpson2017securing, nobakht2016host, sivaraman2015network}.
These approaches usually work by applying cryptographic schemes like identity-based encryption~\cite{jedi} over the existing data flow or introducing new layers like SDN~\cite{demetriou2017hanguard} or auditing services~\cite{trust-but-verify}.

With the full recognition of today's successful cloud-based practice and the emerging efforts to cloud-based security and privacy, we propose to meet the end users' privacy requirements by rethinking smart home realization: can we make the smart home control staying at home?
To be more specific, we aim to build smart homes as independent and self-contained systems solely owned and controlled by end users whose operations do not \emph{rely} on the cloud or other external services.
Our goal is not to prohibit the use of the cloud service; instead, all usages of external resources (\eg remote backup, intensive computation, data retrieval) are authorized and controlled by the local smart home systems.

\begin{figure}[t]
	\centering
	\includegraphics[width=0.45\textwidth]{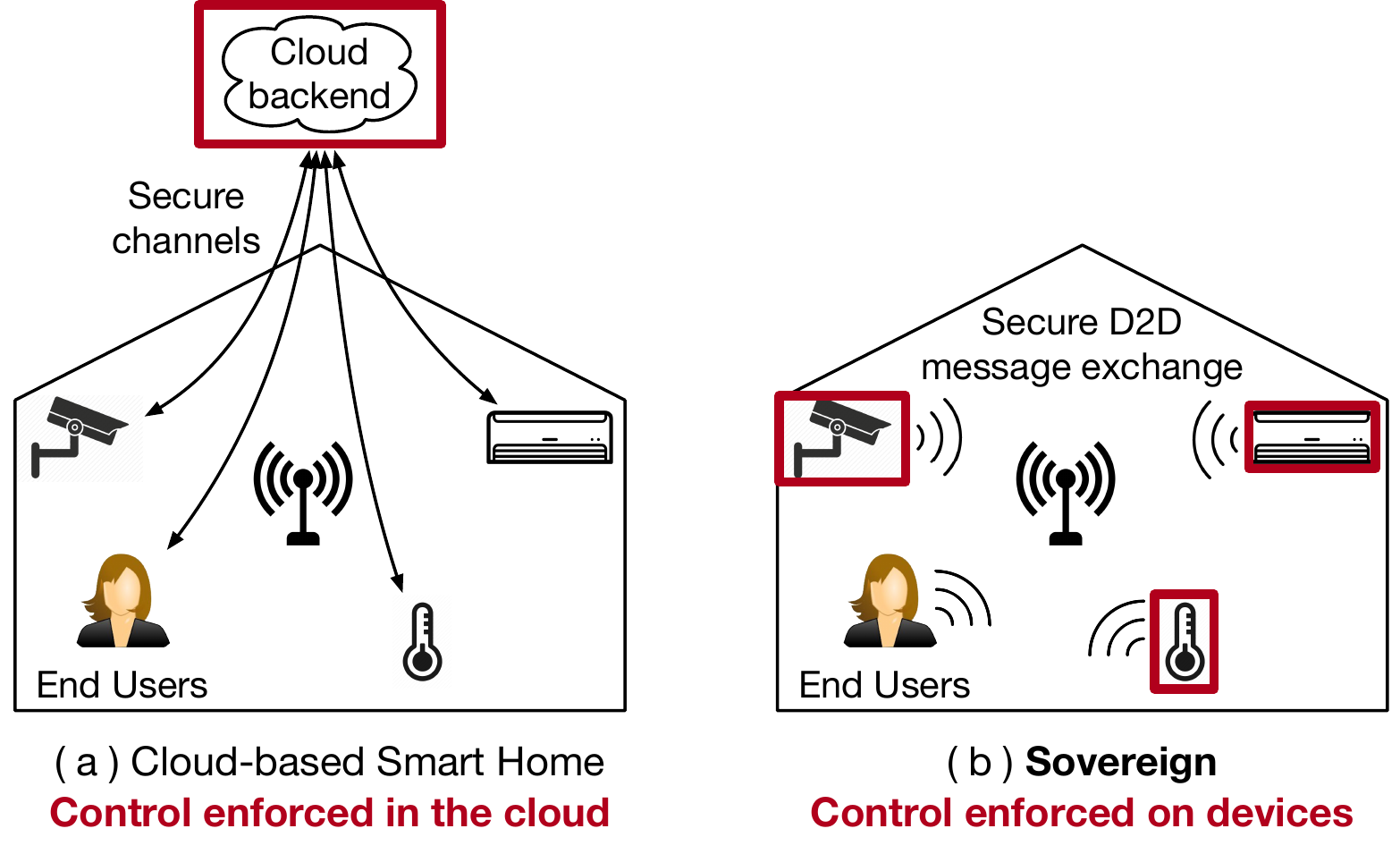}
	\caption{The current practice v.s. \sysname}
	\label{fig:modules_cmp}
\end{figure}

In this paper, we propose \textit{\sysname}---a framework for self-contained smart home systems.
Figure~\ref{fig:modules_cmp} provides a conceptual level comparison between the current practice and \sysname.
\sysname enables each home to act as an autonomous system, where all the  entities belonging to the smart home system (\ie devices and local or remote applications) are managed locally.
To enable reliable and secure communication in the home environment, \sysname takes the approach of data-centric networking and security.
First, every home entity as well as all available resources, \ie content, services, keying material, and security policies, are named with hierarchical and meaningful names.
Second, home entities directly exchange messages using data names over the broadcast wireless network, without going through any centralized message broker.
At the same time, security policies that regulate named entities' access to named resources, are enforced on the smart home devices instead of by a cloud provider.


\sysname is implemented as a software development kit (SDK) that provides developer-friendly application programming interfaces (APIs) to smart home device manufactures and application developers.
Programming a device or an application with the \sysname framework enables it to connect to the localized smart home system and to be controlled by home users without external dependencies.
Users control their homes through user interfaces provided by the smart home applications programmed over \sysname.  The application development, \eg human-computer interaction (HCI), and the home security policy definitions are beyond the system framework, thus not covered in this paper.


%

\mypara{Contributions}
Our contributions are twofold.
First, we design and prototype \sysname, demonstrating an alternative approach to building smart home systems that can fully protect user privacy.
Our prototype offers a proof of evidence that a smart home can indeed be built as an independent and self-contained system, with end-to-end security to enhance system security and direct device-to-device message exchanges to increase resilience and to overcome single points of failure.

Second, we implement \sysname as an open-source and fully operational smart home SDK~\cite{ndn-lite}.
\sysname provides APIs with usability similar to that of today’s practice by offering the same publish-subscribe model, a widely-adopted programming pattern.
In addition, we implement a lightweight Named Data Networking (NDN) protocol stack for the Internet of Things (IoT) providing data centricity while working smoothly with constrained devices.
\sysname follows a modular design and can be adopted by any IoT platform that supports the standard C language.
Along with the development platform, we also provide a python-based prototype controller application to let end users control their homes.
Our evaluation shows that \sysname is developer-friendly and exhibits low network latency, low computation overhead, and a small resource footprint, which allows \sysname to be used on constrained devices.

\mypara{Outline}
The rest of the paper starts with our motivation and \sysname's design principles in \S \ref{sec:motivation}.
We then describe the \sysname framework in \S \ref{sec:overview} and outline its implementation, including \sysname's SDK with usable APIs, in \S \ref{sec:integration}.
We provide the evaluation results in \S \ref{sec:evaluation}, talk about related works in \S\ref{sec:relatedwork}, and give some discussions in \S\ref{sec:discussion}.
Finally, we conclude our work in \S\ref{sec:conclusion}.


\section{Motivation}
\label{sec:motivation}

\vspace{2mm}

\subsection{Cloud-based Smart Home: Your Privacy is in the Cloud}
\label{sec:cloud-based}

Today's dominating IoT frameworks center around the cloud.
Popular cloud-based smart home platforms, such as Samsung's SmartThings~\cite{smartthings} and Amazon's AWS Home IoT~\cite{awsiot}, account for a large market share.
For example, according to Samsung, SmartThings has 120 million active users worldwide by the end of 2020~\cite{smartthings-market-size}.

To understand the role played by the cloud in a smart home system,
we analyze how Samsung SmartThings works as an example.
The SmartThings ecosystem~\cite{smartthingsdoc} has three main components: \first the SmartThings cloud, \second cloud-connected devices, and \third cloud-hosted applications.
SmartThings requires that device vendors and application developers pre-install the SmartThings cloud certificate into their products and register their products to the SmartThings cloud in advance, so that SmartThings can recognize their unique IDs and public keys.
To add a new device to a smart home system, the homeowner performs a simple operation, \eg scanning a QR code or pressing a button on the device. This operation pairs the device with the user's account in the cloud backend,  authenticating the device to the homeowner's smart home system.
As shown in Figure~\ref{fig:modules_cmp}, all devices and applications in a home system establish secure connections, either directly (\eg through Transport Layer Security (TLS)) or indirectly (\eg via a WiFi-connected home hub), to the SmartThings cloud, where they take control commands and communicate with each other.
Importantly, the cloud manages access to all the home entities using OAuth~\cite{oauth}.
An application or device must first obtain OAuth tokens from the cloud before it can access other home resources.

To summarize, the SmartThings cloud backend
\first holds the trust anchor (\ie cloud's certificate), and maintains identities of all the devices and applications in a smart home,
\second serves as a rendezvous point for data exchange among devices and applications; and
\third executes security policies (also called control policies) by controlling access to home resources.
In addition, the cloud is also used as storage for smart home application data in general.
Similar designs are observed in other cloud-based home systems, including AWS Home IoT~\cite{awsiot}, Google Assistant Smart Home~\cite{google-assistant}, and Microsoft's Azure IoT system~\cite{microsoftiot} (see details in Appendix~\ref{appendix:cloud}).

These cloud-based systems expose home users' privacy to the cloud backend:
the cloud can view all user commands and data exchange from the home system\footnote{On today’s home IoT market, Apple’s HomeKit [3] supports local communication among devices and allows users to directly control their home IoT system, reflecting a shared goal with our work. However, HomeKit is not cloud-independent, as it still relies on the cloud for device authentication and identity management.}.
When home application data, such as video camera footage, is stored in the cloud, it adds further privacy concerns~\cite{choe2011living, device-breach, wyze-breach}.
In addition to user privacy concerns, the ``control by cloud'' design increases the overall system complexity when multiple cloud backends get involved. For example, to automate SmartThings devices with IFTTT~\cite{ifttt},
users need to grant IFTTT access right to the SmartThings cloud backend, which introduces additional cross-cloud data exchange and security configuration (\eg through OAuth).
Furthermore, the home systems' dependency on the cloud backends provided by tech giants further intensifies Internet consolidation~\cite{consolidation}.
Finally, although today's cloud services seem reliable in general, the last few years did witness several large-scale cloud service outages, each time disrupting large numbers of users~\cite{cloudOutage1,cloudOutage2}.

We note that there do exist smart home products that are locally controlled and operated. However, these systems lack features, such as secure communication between devices, or exhibit a single-point-of-failure, as we discuss in \S\ref{sec:relatedwork}.

\subsection{Letting Smart Home Stay Home}

Recent years witnessed great successes of cloud computing which takes advantage of economy of scale. Therefore, it is natural that smart home developments latched to the readily available cloud infrastructure for quick development and deployment.
However, when examining the details, controlling home IoT systems from the cloud does not seem to share the same benefits as resource-intensive applications.
Although a home system may need to run computation-intensive apps, \eg face recognition, which can benefit from cloud assistance, the control logic of a smart home does not need heavy computation and could be done locally, especially as smart home devices become cheaper and more powerful over time.
Controlling smart homes from the cloud seems mostly driven by business incentives rather than a technical necessity.

From a technical point of view, bringing the smart home control from the cloud to the home provides the most effective solution to enhance user privacy.
%
To figure out the needed functional support, \S\ref{sec:cloud-based} already identified the cloud's role in smart homes:
\first the cloud hosts a smart home controller;
\second individual home entities communicate with each other by using the cloud as a message broker, and
\third security is provided by the secure connections from individual devices to the cloud using TLS.

Thus the first requirement for building a home-based smart home system is a local controller that takes the user's configuration and manages the home accordingly.
The second requirement is enabling communication between smart devices and applications (hereafter \emph{entities}).
Lastly, after moving the control out of the cloud, smart homes must be protected from local surrounding adversaries, \ie one must add strong security protection to local communications in the smart home system.

Addressing the above-mentioned requirements brings up the following identified challenges.
Setting up a controller at home differs from running a controller in the cloud in at least two ways: the controller needs network and security configurations, and the controller can be a single point of failure; neither is an issue when the controller is in the cloud.
Communication between individual home entities, on the other hand, can be trivially achieved by supporting direct device-to-device (D2D) communication, assuming each home is connected by a WiFi wireless network that is broadcast in nature.
Accordingly, security must be applied directly to D2D communication.

\subsection{Design Choices}
\label{sec:bkg:choice}

We identified mitigating the single point of failure and providing secure D2D communication as the challenges in realizing a self-contained home system.
The first one includes two cases: the controller fails or gets compromised. One can mitigate the failure by providing home entities the ability for secure message exchange without involving the controller; mitigating the compromise can also be achieved, which we discuss in \S\ref{sec:discussion}.
To achieve secure D2D communication, we see the following two possibilities: \first a channel-based solution, \eg utilizing TLS, or \second a data-centric solution, as provided by Information Centric Networking (ICN).

\begin{figure}[t]
	\centering
	\includegraphics[width=0.4\textwidth]{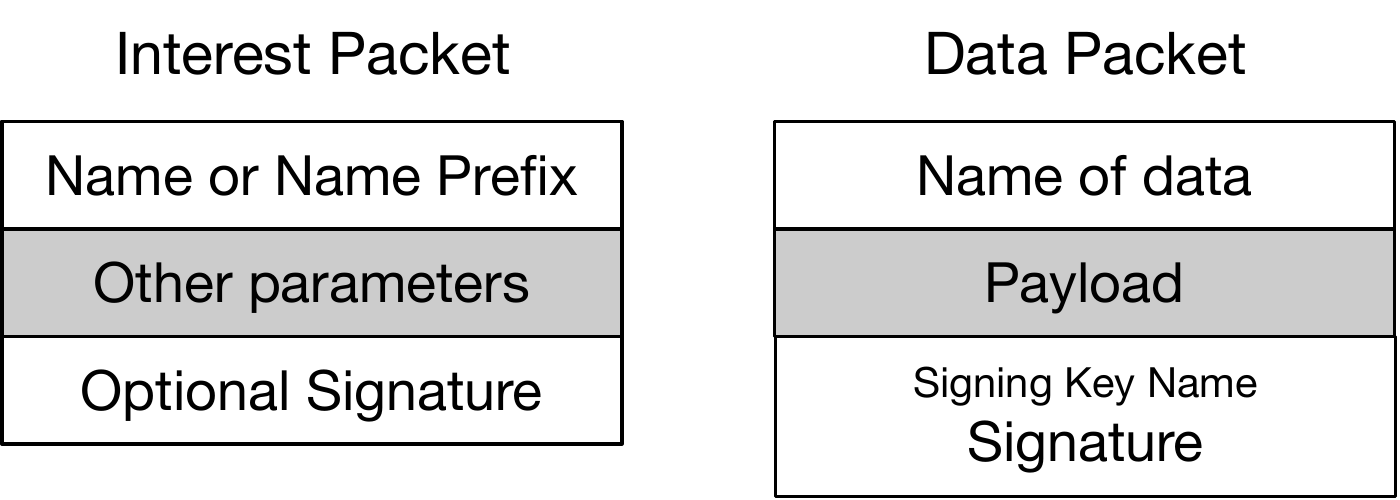}
	\caption{Interest and Data packets in NDN}
	\label{fig:ndn-pkts}
\end{figure}

Channel-based communication and its security model has been the default choice in network system design for many years.
However, it is not necessarily the best fit for all use-cases.
Deploying channel-based security in a smart home scenario requires all home entities to set up bidirectional TCP/IP connections, secured using TLS, which introduces limitations and complexity in various aspects.
First, TLS does not utilize the broadcast/multicast capabilities of wireless home networks.
For example, to turn on all the lights at home, a simple way would be to issue a verifiable command through multicast.
However, with TLS, the command must be sent through each TLS channel to all the lights.
TCP/IP connections also require a mapping between application-level identities and network identifiers, \ie IP addresses, and
maintaining such mappings requires additional synchronization services.
Moreover, executing control policies (\eg the air conditioner can access temperature data) requires application semantics, which must be either installed at all the entities or otherwise has to be performed by the controller.



In contrast, data-centric networking and its security model provide a new option:
Each piece of data is identified by a semantically meaningful name, and security is carried in the data instead of on the channel.
This data-centric networking model is realized by the Named Data Networking (NDN) architecture.
We introduce the key function of NDN below to prepare readers ready with data-centric security in the next section.

In NDN, applications pull data from the network by requesting it using the data's name.
A request for data, called \textit{Interest} packet, carries the name of the desired Data. Packets carrying the fetched data are called \textit{Data} packets.
When produced, each Data packet is secured by the producer application by adding a digital signature generated by the producer's private key.
The signature allows data consumers to verify the authenticity of every received Data packet, so that a Data packet can be retrieved not only from the producer but also from storage components or in-network caches.
NDN directly uses application layer names for data fetching, where the names are semantically meaningful and follow a hierarchic structure in general, similar to URLs used in today's Internet.
As Figure~\ref{fig:ndn-pkts} shows, both NDN packet types carry data names; Data packets also carry the requested payload and a cryptographic signature.
To facilitate data retrieval, applications make use of \emph{naming conventions}, \ie well-established naming patterns, to help consumers construct names of desired data.
As an example naming convention, the temperature produced in the bedroom of Alice's home can be named as \ndnname{/alice/temp/bedroom}. This allows a potential requester to fetch data without having to discover the name first.

The properties of NDN (and ICN in general) enable secure D2D communications by securing data directly.
In a home network, an entity can directly fetch desired data by sending Interest packets to the local broadcast network, eliminating the additional layer of indirection between IP address and application layer data identifiers in a TCP/IP network.
Security is realized in a data-centric way by letting the data producers sign and encrypt the data using their keys.
In this way, a piece of secured data can be fetched and verified by multiple entities while only those who have sufficient keys can access the payload.
The semantically meaningful name carried in each message can be directly used to construct and enforce security policies (see \S \ref{sec:sovereignnamedesign} for an example).

\section{Design of \sysname}
\label{sec:overview}

In this section, we first use an example to provide an overview of \sysname's system design, then introduce individual components from an entity's perspective.

\subsection{An Overview}

\begin{figure}[t]
	\centering
	\includegraphics[width=0.85\linewidth]{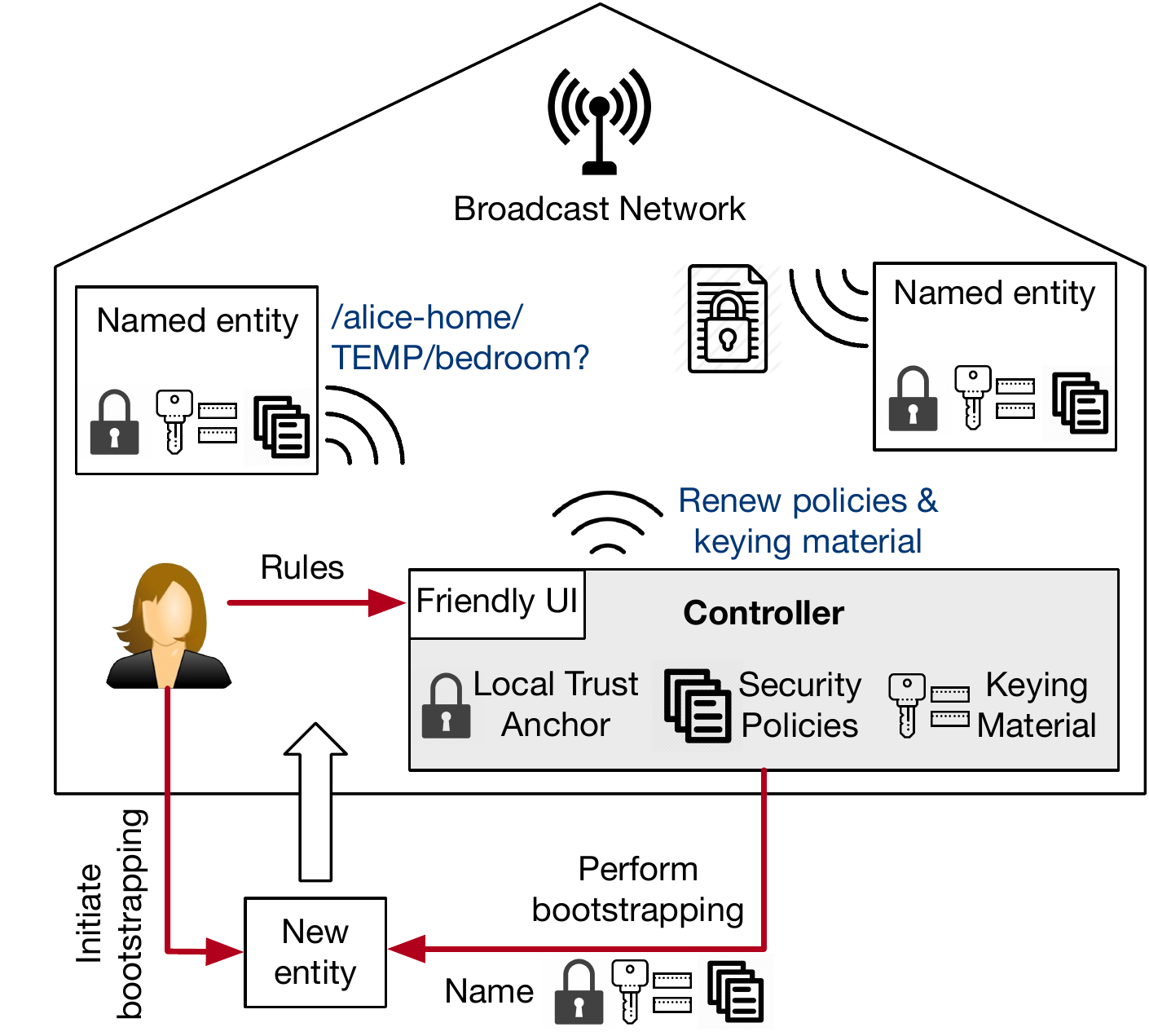}
	\caption{High-level overview of the \sysname Home System}
	\label{fig:overview}
\end{figure}

As stated, today's cloud-based smart home design puts trust, control, and security to the cloud.
In \sysname, this is fundamentally changed by bringing the trust anchor to a local controller.
Both networking and security in \sysname center around names:
\begin{itemize} [leftmargin=*]
\item Content fetching is implemented by sending requests carrying the content name and service invocation is realized by issuing commands carrying the service name.

\item Security is implemented by expressing constraints on different names. For example, service access control is to constrain which named entities can issue commands to which named services.
\end{itemize}
Individual entities enforce security policies by directly securing the D2D communication, and \sysname removes the single rendezvous point since the data-centric communication takes place directly on the local broadcast network.

Our new framework impacts the way of networking and security, but is not supposed to affect the application logic, and hence, does not change end users' experiences of using smart home compared with today's cloud-based systems:
A home user Alice only does minimal operations to set up the whole system and adding new entities, \eg by scanning a QR code mounted on a device with the controller.
Importantly, Alice controls her home by deciding the rules of the system through user interfaces (UIs) provided by the local controller\footnote{We acknowledge that correctly configuring security policies possibly goes beyond the capabilities of most homeowners. However, the support of cleverly designed user interfaces (not in the scope of this work) can ease the configuration process, \eg by providing default policies with best practices.}.

Figure~\ref{fig:overview} visualizes the high-level overview of \sysname.
Let us assume the user Alice installs a new air conditioner (AC) in her home and wants to connect it to \sysname. Therefore, Alice uses the controller to scan a QR code mounted on the AC, which initiates the entity bootstrapping process (see \S~\ref{sec:sovereignsecuritydesign}).
In this process, the device registers itself to the controller and the controller \first installs the system trust anchor to the device, \second sends keying material and security policies needed by the device, as well as \third assigns a name to the device according to the naming conventions (see \S~\ref{sec:sovereignnamedesign}).
In \sysname, the security policies are based on names and are converted from user-decided rules by the controller.
In addition, the entity's name is associated with a public key pair whose public key will be certified by the controller, and the private key is kept secret by the AC.

\begin{table*}[t!]
	\newcolumntype{C}{>{\raggedright\let\newline\\\arraybackslash\hspace{0pt}}m{0.3\linewidth} }
	\newcolumntype{D}{>{\raggedright\arraybackslash} m{0.6\linewidth} }
	\footnotesize
	\resizebox{\textwidth}{!}{
		\begin{tabular}{CD}
			\toprule
			\textbf{Type} & \textbf{Naming Convention} \\ \midrule
			Device\&Application
			& /$<$home-prefix$>$/$<$service$>$/$<$location$>$/$<$entity-id$>$*
			\newline {\footnotesize \eg /alice-home/AirCon/bedroom/north-ac-1} \\
%

			\midrule

		    Commands to executables
			& /$<$home-prefix$>$/$<$service$>$/$<$\textbf{scope}$>$/CMD/$<$cmd-id$>$**
			\newline {\footnotesize \eg /alice-home/AirCon/\textbf{bedroom/north-ac-1}/CMD/set-temp -- device-level command}
			\newline {\footnotesize \eg  /alice-home/AirCon/\textbf{bedroom}/CMD/set-temp -- room-level command}
			\newline {\footnotesize \eg /alice-home/AirCon\textbf{/}CMD/set-temp -- home-level command}
			\\
			\hline

			Service's Content
			& /$<$home-prefix$>$/$<$service$>$/CONTENT/$<$location$>$/$<$entity-id$>$/$<$content-id$>$**
			\newline {\footnotesize \eg /alice-home/TEMP/CONTENT/bedroom/senor-1/temp}
			\\
			\hline

			Encryption/Decryption Key
			&  /$<$home-prefix$>$/$<$scope$>$/EKEY \newline  /$<$home-prefix$>$/$<$scope$>$/DKEY\\

			\midrule

			Security Policy
			& /$<$home-prefix$>$/RULE/$<$location$>$/$<$entity-id$>$** \\
			\bottomrule
		\end{tabular}
	}

	{\raggedright
		\footnotesize
		Notation: A component with $<>$ represent a variable. A component without $<>$ represent a constant string component. \\
		$*$: Actual service command, content, and policy NDN Data packets will have a timestamp suffix to achieve the data uniqueness. \\
		$**$: The corresponding identity key name is the identity name with a ``KEY/$<$key-id$>$'' suffix. \par
	}
	\caption{Naming Conventions of entities and data (\ie services, content, and keying material)}
	\label{tab:naming_convention}

\end{table*}

After joining the system, the AC starts running by fetching the temperature data and setting the target temperature accordingly.
In addition, the AC will also close bedroom windows when it is running.
In \sysname, the temperature data fetching and window command sending all take place directly between the involved devices over the broadcast network in the form of NDN Interest--Data exchange.
To get temperature data, the entity directly broadcasts the desired data name so that any other entities who have the data can reply.
After fetching the Data packet carrying the temperature, the AC will first verify the packet is signed by another home entity, whose certificate is endorsed by the trust anchor.
Then, to access the payload, the AC needs to decrypt it using keys obtained from the controller.
When issuing a command to the window, the AC will also name this command according to the naming convention, encrypt the payload, and sign it with its own private key.
Therefore, windows in the system can fetch the command by the name.
If the AC is allowed by the security policies to generate such a command, home windows will successfully verify, decrypt, and execute the command.
A detailed description of these workflows is provided in Section~\ref{sec:sovereignworkflows}.

\subsection{Name Design}
\label{sec:sovereignnamedesign}


One key difference of \sysname compared to IP-based systems is the use of semantic names in the network. Since endpoint addresses are used for decades, the advantage of name-based communication might not be obvious. This is why we first focus on answering the question: \textit{Why bothering with designing semantic naming schemes?}
%
%
First, semantic metadata is needed to represent identities and data when realizing application logic.
In \sysname, naming conventions allow entities to follow established rules to infer the name of the desired data.
Second, it is possible to design security policies based on names. This means that security policies can directly define which \emph{named} entities are allowed to produce or access which \emph{named} resources.
Hence, in \sysname, name-based security policies allow inferring which entities are granted access to certain resources when looking at a Data packet's name.


\mypara{Naming Conventions}
Prior studies~\cite{he2018rethinking, zeng2019understanding} suggest that smart home access control and authentication systems should be flexible enough to support a wide variety of use cases and types of relationships that exist in homes.
Therefore, \sysname's naming conventions embed the attributes held by entities and data into semantic names. These attributes include an identifier of the home system, a service type, the location of the corresponding entity or data, and a resource type.
Table~\ref{tab:naming_convention} summarizes the structure of names and naming conventions used in \sysname.
The home prefix is defined during the system setup and better to be globally unique considering potential communication among homes in a neighborhood.
One approach is by attaching a random string to a user-specified name.

\sysname allows flexible service invocation and content fetching by constructing different names based on the application's need.
For indicating the capabilities of \sysname's naming scheme, we provide a small example of an AC in the bedroom of Alice's home.
The AC listens to requests under three prefixes that allow other entities to send control commands on a device-level, room-level, and house-level.
\begin{center}
\footnotesize
\raggedright
~~~~/alice-home/AirCon/\textbf{bedroom/north-ac-1}/CMD/set-temp \\
~~~~/alice-home/AirCon/\textbf{bedroom}/CMD/set-temp \\
~~~~/alice-home/AirCon/CMD/set-temp \\
\end{center}
Thus, the controller or other authorized entities can flexibly control the home temperature in the desired granularity.
Importantly, semantic naming combined with broadcast media allows sending one command to control multiple devices.
This is a sharp contrast to IP-based systems, where commands address single devices only.

\mypara{Name-based Security Policies}
\sysname's naming conventions facilitate the use of name-based security policies.
As elaborated earlier, security policies generally specify entities that are authorized to perform certain operations.
In \sysname, flexible control can be achieved by specifying the names of entities and resources in the policies.
Given the name $P$ representing one or multiple entities, and the name $R$ representing one or multiple resources. One can limit $P$'s authorized actions to $R$ by defining a security policy.
Such a policy can be written as a triple based on $P$ and $R$, which can be either be specific names, name prefixes, or regular expressions of names.
\begin{center}
	\emph{$<$$P$'s name, verb, $R$'s name$>$}.
\end{center}
For example, the verbalized policy ``\textit{the controller can command all door locks}'' can be written as the name-based policy
\textit{$<$controller name, produce, door lock command prefix$>$}.
As another example, the verbalized policy ``\textit{temperature sensors can produce temperature data}'' can be represented by the policy \textit{$<$prefix of temperature sensors, produce, temperature content prefix$>$}.

We want to note that using names in security policies is assumed to be secure since names are assigned and certified with the entity's certificate. This certificate is derived from the trust anchor in the entity bootstrapping phase (see \S\ref{sec:sovereignsecuritydesign}).
Thereby it is ensured that entities can not spoof the names of other entities.



\subsection{Security Design}
\label{sec:sovereignsecuritydesign}

%

In \sysname, the authentication and access control are managed by the controller but enforced distributedly in D2D communication.
In general, after converting user-defined rules into name-based security policies, the controller distributes these policies among all entities.
During runtime, all Data packets are signed and encrypted by producers, and verified and decrypted by consumers.
In the verification step, consumers consult the available security policies and verify whether data producers are authorized to sign the piece of data before consuming its content or executing the command.

\mypara{System Setup}
In the system setup phase, the controller generates an asymmetric key pair.
The public key is bound to the home prefix and published as a self-signed certificate. This certificate represents the home's trust anchor.
The trust anchor will be installed on each entity during the entity bootstrapping.
The private key is kept secret by the controller and is used to sign security policies, cryptographic keying material, and certificates for new entities.

\mypara{Entity Bootstrapping}
In the entity bootstrapping phase, a new entity joins the system and learns cryptographic keying material and security policies required for later communication.



The process takes place in the default broadcast media and starts with mutual authentication between the controller and the entity, in which out-of-band operations may be needed (\eg QR code scanning).
Then, the following information is secretly exchanged:

\begin{itemize} [leftmargin=*]
	\item The trust anchor certificate is installed on the new entity.

	\item The controller assigns the new entity an appropriate name according to the naming convention (supported by additional input from the homeowner, \eg entity location or customized device name).

	\item The controller issues a public key certificate binding the entity's public key and name together.
	The certificate is signed with the private key of the trust anchor.

	\item The entity will obtain security policies and symmetric cryptographic keys that the entity is authorized to have.
	These keys will be used in data encryption and decryption for access control and privacy protection purposes.
	In addition, the new entity and the controller will negotiate another symmetric to encrypt the future cryptographic material issued by the controller to this entity.

\end{itemize}
An implementation of \sysname's entity bootstrapping process is discussed in~\cite{ssp-li}.

%
%
%
%
%

Bootstrapping hardware-independent applications follows similar steps as given above.
Note that applications can run either locally (\eg software executed on the home hub) or remotely (\eg software hosted in the cloud).

\mypara{Enforcing Security Policies}
While name-based security policies are maintained by the controller, individual participants enforce these security policies for the D2D communication.
To elaborate on how these policies are enforced, we differentiate between two types of security policies: \first \textit{produce}-policies define which entities are allowed to produce data of a specific name pattern (\eg sending an actuating command to a door lock),
\second \textit{decrypt}-policies define the entities that are allowed to access data of a specific name pattern (\eg data from specific sensors).

\textit{Produce}-policies are enforced by data receivers.
To be more specific, after authenticating a Data packet, the receiving entity extracts the data name and producer's name from NDN Data's name and signature fields.
These names allow checking whether the data was generated by an authorized entity defined in security policies.
For example, the following policy defines that temperature content can only be signed by a temperature sensor, where \ndnname{/alice-home/TEMP} is the shared prefix of all home temperature sensors, and \ndnname{/alice-home/TEMP/CONTENT} is the common prefix of temperature content.
\begin{center}
	\ndnname{$<$/alice-home/TEMP, produce, /alice-home/TEMP/CONTENT$>$}
\end{center}
Combining the Data's name, the producer's identity, and the available security policies allow receiving entities to reject temperature content produced by unauthorized parties.

The same procedure is applied for restricting entities to issue commands.
For example, a \textit{produce}-policy as follows indicates that all the automation applications running on the home hub named ``hub-1'' can invoke executables whose names match the specified regular expression\footnote{The regular expression \ndnname{$<>$*} matches zero or more name components.}.
\begin{center}
\ndnname{$<$/alice-home/AUTO/hub-1, produce, /alice-home/LOCK/$<>$*/CMD$>$}
\end{center}
Before executing the issued command, the receiving entities first check the verified producer name against the available security policies and rejects the command when issued by unauthorized entities.

\textit{Decrypt}-policies are enforced by utilizing data encryption.
To be more specific, every entity can fetch data by emitting an Interest packet carrying the desired content name.
The encrypted data, however, can only be accessed when having access to the correct decryption key.
In \sysname, the controller is providing all encryption and decryption keys, and allows authorized entities to obtain certain keys%
\footnote{The \sysname controller makes encryption and decryption keys available for authorized entities. After the distribution of the keying material, the controller is not involved in the D2D communication, allowing entities communicating securely without requiring a central message broker.}.
This reduces access control to maintaining the access of corresponding decryption keys.
For example, the \textit{decrypt}-policy ``bedroom AC can read the temperature'' is written as follows:
\begin{center}
\ndnname{$<$/alice-home/AirCon/bedroom, decrypt, /alice-home/TEMP/DKEY$>$}
\end{center}
The subject name \ndnname{/alice-home/TEMP/DKEY} represents the decryption key to the content produced under the temperature service.


\mypara{Privacy Protection}
\sysname protects the privacy carried in network packets with three methods:
\first As stated, the payload of all Data packets is only sent in encrypted form.
\second Since semantic names may also leak sensitive information, \sysname applies name obfuscation.
That is, instead of using plaintext name components, \sysname can choose to use pseudonyms derived from a keyed hash (\eg HMAC of the name component with a random key generated by the controller).
Besides, service-specific name components, such as \name{content-id} and \name{command-id} are encrypted with the payload's encryption key.
\third In \sysname, the data exchange happening in the local network is invisible to the outside.
When homeowners allow services provided by trusted remote entities, the latter can only access the data allowed by the security policies.

\mypara{Key Distribution}
As stated, encryption and decryption keys are first distributed during the entity bootstrapping process.
In the system runtime, since each key has a lifetime and security policies can change (\eg end users add or revoke access rights), keys need to be periodically renewed.
Therefore, entities will need to retrieve the updated keys from the controller.
In \sysname, decryption keys will be encrypted for each authorized entity using the shared symmetric key between the entity and the controller, and they can be retrieved by the key name (Table~\ref{tab:naming_convention}) with the entity's name as the suffix.
Importantly, this process can be asynchronous: the controller can pre-provision Data packets carrying the encrypted renewed keys into a storage component, allowing the key renewal process to operate when the controller is temporarily down.

\subsection{Putting Everything Together}
\label{sec:sovereignworkflows}


The previous sections explained individual components of \sysname. In this section, we show how everything works together to build a local, user-controlled smart home system.

Reviewing current smart home systems shows that the individual entities can be classified into two groups.
The first group represents devices that provide executables.
These devices can be as simple as a light bulb that can be switched on or off, but also more security-critical, such as the door lock of the front door.
The second group represents devices that produce content., \eg sensors that monitor the smart-home environment, providing temperature information or motion detection.
Also, some devices combine both groups.

When invoking an executable, or when requesting content, entities are typically not interested in communication with a specific device or application, but in a specific service on a certain location (\eg switching on the kitchen light, getting the ambient temperature of the living room).
Without knowing the actual entity required for the action, the name can be inferred by joining the service name with the intended location scope (\eg \ndnname{/alice-home/Light/kitchen/CMD/switch-on}, \ndnname{/alice-home/TEMP/CONTENT/living room}).
When it comes to networking, we differentiate between actuating messages used for invoking executables, and content consumption messages for reading data.
To fetch a piece of content, the desired content name is encoded into an Interest packet for broadcast.
In contrast, when issuing a command, the command is encoded in a Data packet to carry the signature and sufficient parameters in the payload.
The command sender will emit a notification Interest to the broadcast network so that relevant other entities can fetch the command.
Note that command fetching only needs to take place once because other entities can also hear the command via broadcast.

To fulfill security policies, the produced content and commands are encrypted and signed by its producer.
Hence, when receiving a Data packet, the receiver can first verify whether the producer is authorized to produce the data by checking the signature against security policies.
In case the producer is authorized, only authorized receivers can access the decryption key and successfully decrypt the Data packets content. Hence, security policies are enforced.

\section{Implementation of A Usable Framework}
\label{sec:integration}

The \sysname framework is implemented as a smart home Software Development Kit (SDK). Our SDK supports developers and device manufacturers in building \sysname-compatible devices and applications.
The design of our API design aims to provide a high developer-friendliness and is outlined in \S \ref{subsec:integration_pubsub}.
Our SDK's main components are discussed in \S \ref{sec:implementation}.

We also provide a proof-of-concept controller with preliminary UIs.
This controller allows home users to bootstrap devices and applications and to define security policies.
More sophisticated UIs can be designed to allow end users to use \sysname the same way as current smart home systems.
\sysname's main contribution -- enabling a self-contained, local home smart home -- is beneath the applications and should be transparent to end users.

%

\subsection{Encapsulation of Framework Details}
\label{subsec:integration_pubsub}

The core implementation idea of \sysname is to use a publish-subscribe (pub/sub) communication module that encapsulates naming, security, and networking primitives in one API.
Pub/sub is a messaging pattern that categorizes messages into semantically meaningful topics and is often seen in the context of IoT.
Message producers (called publishers) publish messages to topics without knowing the set of message consumers (called subscribers).
Subscribers choose to receive messages under pre-defined topics without the need to know the actual message producers.
Pub/sub is used for two main reasons.
First, pub/sub is data-centric which matches the data-centric networking and security design of \sysname.
Second, pub/sub has been widely adopted in existing IoT frameworks, and hence, it provides convenience for developers.

Pub/sub API is directly built over \sysname's D2D communication as presented in the previous section by handling name prefixes as topic identifiers.
That is, subscribers use name prefixes to decide whether a message is under a certain topic or not.
For example, a message carrying temperature data of the bedroom is mapped to the name prefix \ndnname{/alice-home/TEMP/CONTENT/bedroom}.
This name prefix is further used as the pub/sub topic identifier.
In this way, producers that publish content or commands under a topic is to generate Data packets named under the corresponding prefix.
Subscribing to a topic is implemented by issuing Interests containing the topic's name prefix to fetch relevant data.

\begin{figure}[t]
	\centering
	\includegraphics[width=0.47\textwidth]{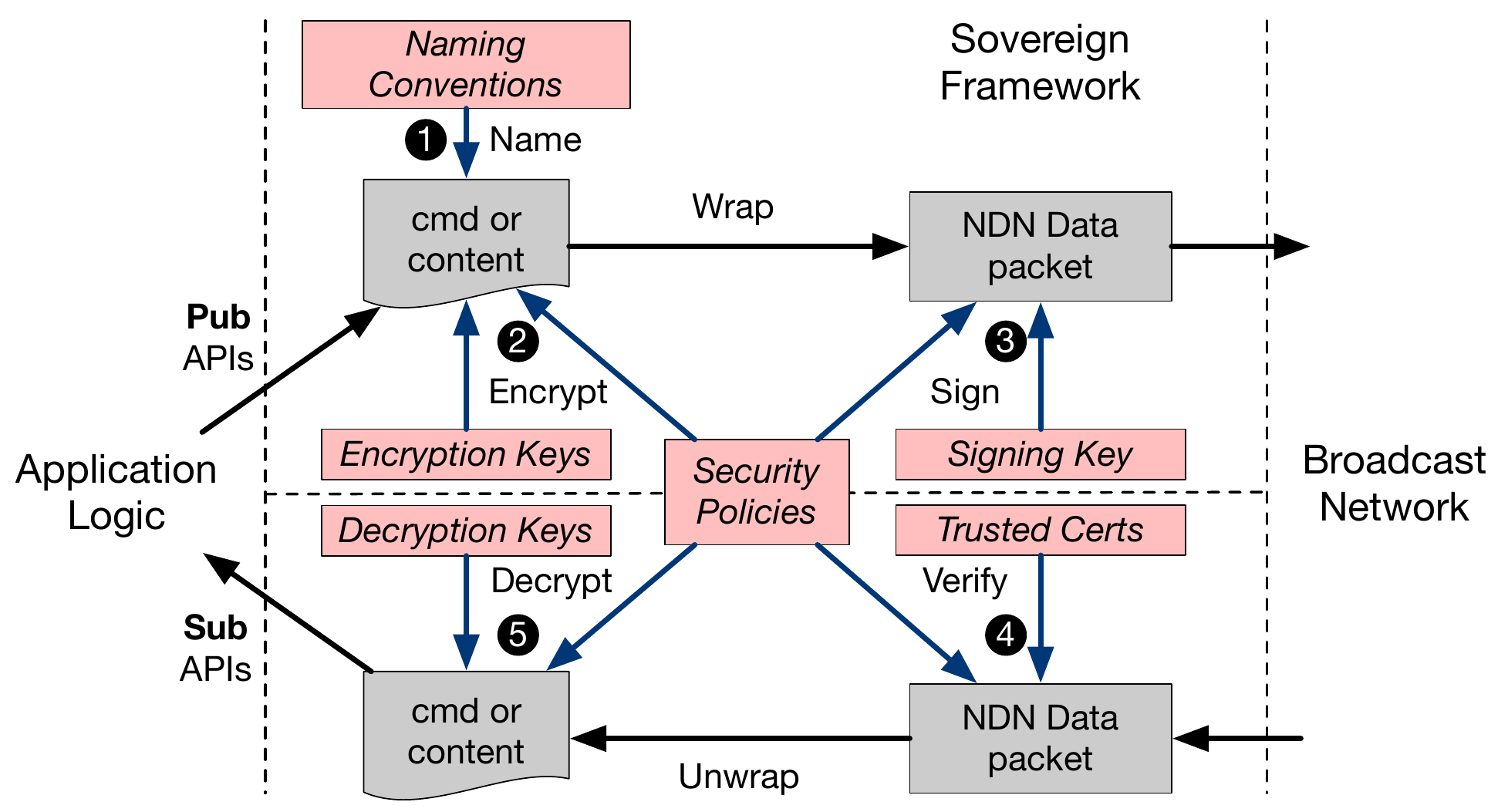}
	\caption{Workflow beneath the pub/sub API}
	\label{fig:modules}
\end{figure}

As indicated in Figure~\ref{fig:modules}, in \sysname implementation, the pub/sub API embeds naming, security, and networking considerations to make them transparent to developers.
We illustrate the underlying workflow with an example, where an application issues a command to set the bedroom temperature to 70\textdegree F.
After calling the publish API, \sysname defines the command name based on application parameters following naming conventions (\ding{202}).
\sysname identifies an encryption key according to the topic and encrypts the payload (\ding{203}). Further, the API wraps the command name and the encrypted payload into a Data packet.
Finally, the API uses the application's private identity key to sign the Data (\ding{204}) and makes it available on the local network.
On the receiving end, after subscribing to a topic, the API automatically fetches the relevant Data packet.
Once a Data packet is received, \sysname verifies its signature and checks against security policies (\ding{205}).
In the next step, the Data is decrypted with the corresponding decryption key ({\ding{206}).
Once verified and decrypted, the command is delivered to the application.
For clarity, we omit low-level protocol details of \sysname's pub/sub transport in this publication. We provide a discussion of the pub/sub design, protocol details, and implementation in a supplemental technical report~\cite{tr-pubsub}.

\subsection{\sysname Software Development Kit}
\label{sec:implementation}

\begin{figure}[t]
	\centering
	\includegraphics[width=0.45\textwidth]{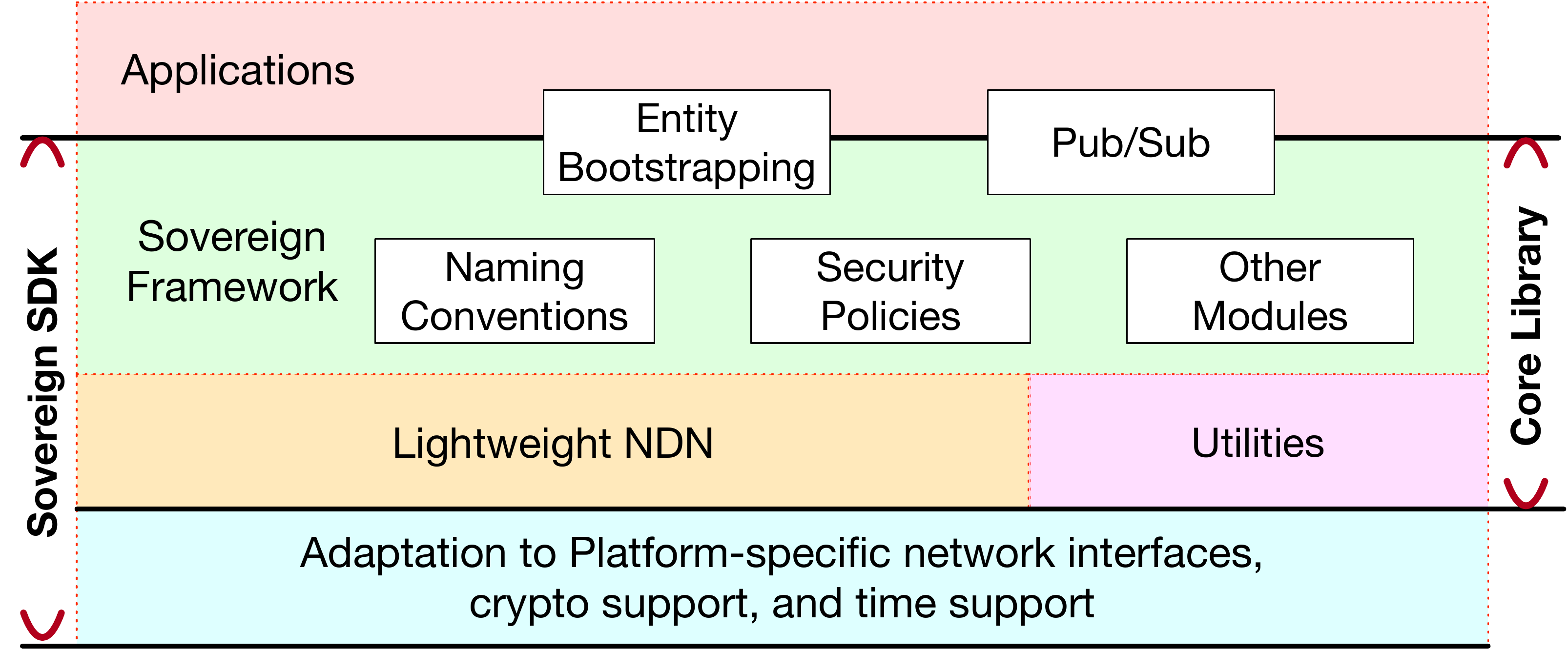}
	\caption{Structure of \sysname's Smart Home SDK}
	\label{fig:ndn-lite}
\end{figure}

\sysname is made available as an open-source, cross-platform software development kit (SDK) built on the C programming language~\cite{ndn-lite}.
The SDK's structure is visualized in Figure~\ref{fig:ndn-lite}.
The core library implements the \sysname framework as detailed in \S \ref{sec:overview}.
As indicated, the only components exposed to developers are the \textit{Entity Bootstrapping} and the \textit{Pub/Sub} APIs.
Other components are transparent for developers.
Moreover, the SDK includes an adaptation layer making the core library work across different platforms and communication media. So far, the adaptation layer is tested for platforms including Linux/Unix, RIOT OS~\cite{baccelli2013riot}, and Nordic NRF boards~\cite{nordic-board}.
Regarding connectivity, the adaptation layer allows using Bluetooth, IEEE 802.15.4, and the legacy TCP/UDP used as link layer protocols.

Also, the \sysname SDK includes a standalone NDN stack that follows the official NDN specification~\cite{ndn-spec-03} and is lightweight enough for constrained devices.
NDN-LITE is required since the official NDN library and forwarder --- ndn-cxx~\cite{ndn-cxx} and NFD~\cite{nfd} --- are not designed for being used on constrained IoT devices.

To be used in real smart home products, device vendors and application developers need to program with \sysname's SDK and write the software together with all dependencies to the target hardware platform (\ie the program memory on the microcontroller).

\section{Evaluation}
\label{sec:evaluation}

\noindent We evaluate the \sysname prototype and answer the following questions:
\begin{enumerate} [leftmargin=*]
\item Can \sysname enhance smart home security and privacy compared with existing cloud-based smart home systems?

\item Is \sysname resilient enough to operate when the controller is temporarily down?

\item Does the use of cryptographic operations on smart home entities impair the usability of \sysname in real deployment, especially on low-power devices?

\item Are \sysname SDK's programming interfaces friendly to IoT developers?
\end{enumerate}


\subsection{Privacy and Security Assessment: A Case Study}
\label{subsec:eval_security}

\begin{figure}[t]
	\centering
    \includegraphics[width=0.5\textwidth]{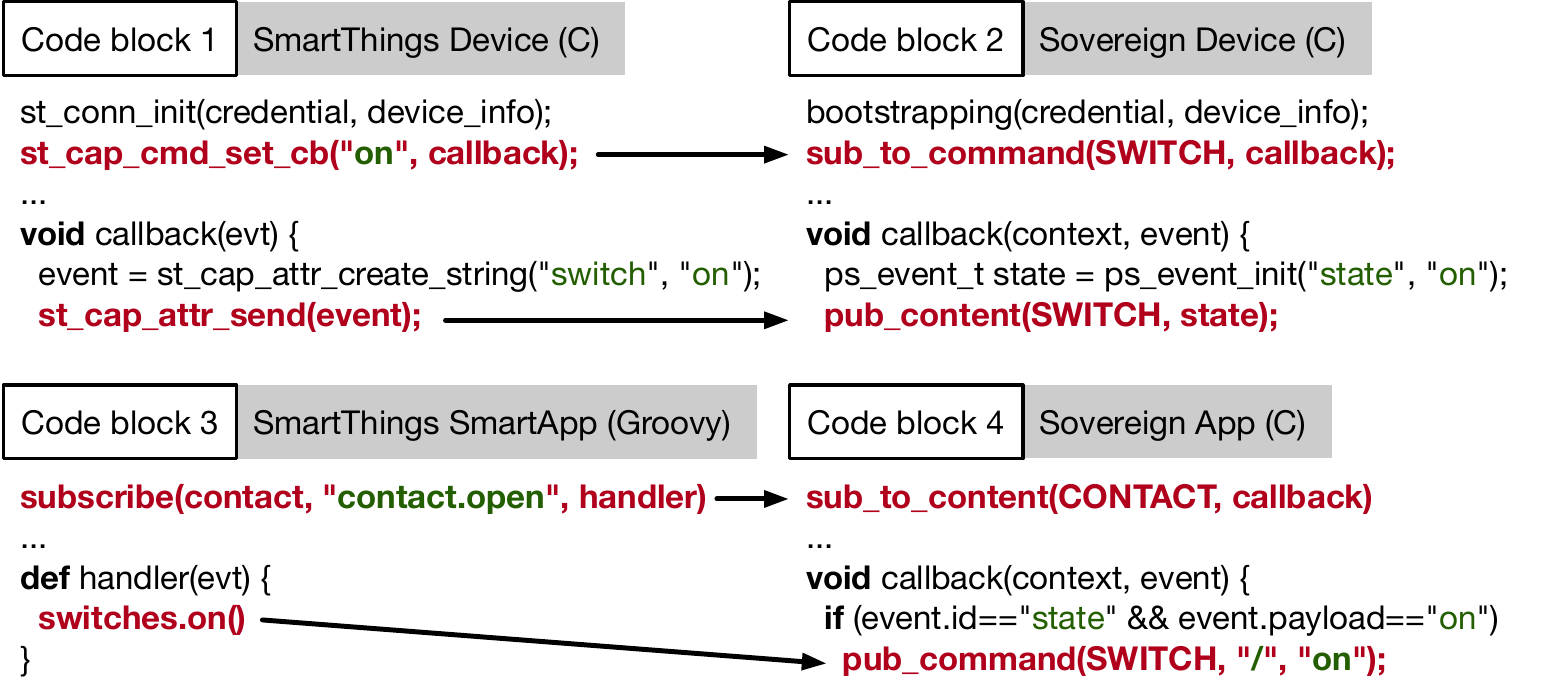}
	\caption{Code snippets in SmartThings and \sysname}
	\label{fig:code-example}
\end{figure}

We first assess \sysname's privacy and security by analyzing packet flows in two real-world applications.
We select two simple yet representative open-source applications~\cite{smartthings-st-device-sdk, turn-it-on-when-it-opens} from the official SmartThings GitHub repositories~\cite{smartthingscommunity}, and realize the same functionality in \sysname.
The first application is designed for a smart switch and the second is an automation applet turning on a switch when a contact sensor is touched. 
The code snippets of the original programs (code block 1 \& 3) and their \sysname-based equivalents (code block 2 \& 4) are compared in Figure~\ref{fig:code-example}.
The switch device application changes its own state to ``on'' when it gets a turn-on command, and the automation applet subscribes to the state of a contact sensor and turns on the switch when the contact sensor is touched.

In the SmartThings device application (code block 1), the first line of the code securely connects the device to the home's cloud-backend.
After that, the cloud recognizes the new device, learns its profile, and registers it to the device database for the home on the cloud.
In contrast, the first line of the \sysname device application (code block 2) bootstraps the device to a local controller. All sensitive information that is transmitted in the bootstrapping phase stays in the local network and is protected by encryption.

The first line of the automation applet of the SmartThings (code block 3) and the \sysname application (code block 4) subscribes to a given topic.
However, the underlying operations differ: the SmartThings application notifies the cloud backend about its interest in the given service, while the \sysname application starts listening to data published under the given name prefix in the home network.
Similarly, when the SmartThings application turns on the switch, the command is sent to the cloud backend, where the command is verified and sent back to the switch at home.
In contrast, publishing a command in \sysname means producing a new Data packet and making it accessible in the local home network.

Reflecting the above comparison, we see that \sysname provides the same functionality as cloud-based smart homes via local secure communication.
The home activities and data are not accessible to the cloud unless users explicitly grant access rights to a cloud service.

\subsection{Resilience, Performance, and Overhead Evaluation}
\label{subsec:eval_resilience}

We conduct our evaluations using an Intel Core~i7 laptop, a Raspberry Pi 3B (RPI), and an nRF52840 board~\cite{nordic-board} to mimic smart home entities with different capabilities.
RPIs are widely used in IoT projects and are often used for prototyping. Those boards are equipped with an ARM Cortex A53 @1.4GHz processor and 1GB RAM.
We classify the evaluated nRF52840 chip with its 32-bit Cortex M4@64MHz CPU, 1MB ROM, and 0.25MB RAM%
\footnote{We acknowledge that a part of current IoT devices shows lower capabilities. However, market studies~\cite{marketwatchMCU, icinsight-report-2015, icinsight-report-2018, icinsight-report-2019, eetimes-turn-32bit} have seen the market turning to 32-bit microcontrollers. Hence, we assume the nRF52840 as an appropriate candidate for future smart home systems.}
as constrained hardware.
The experiments with RPI are over WiFi connectivity and use a laptop equipped with a Core~i7 processor as the \sysname controller.
Evaluations involving the nRF52840 are conducted over IEEE 802.15.4 and use a simplified controller installed on another nRF52840.

\mypara{Resilience to Controller Outage}
To evaluate \sysname in the case of temporary controller outage, we first set up the system and bootstrap new devices with the controller.
After some time, we bring the controller offline and test whether the content fetching and service invocation can still be finished and whether the security policies are still executed as normal.

The experiments presented below confirm that the communication happens in a D2D manner and is secured without a controller involved, showing \sysname's ability to continue operation even when the controller is temporarily down.


\mypara{Latency Comparison with AWS IoT}
We first evaluate the latency of the common operations, including \textit{entity bootstrapping}, \textit{content delivery}, and \textit{command delivery}, between AWS IoT and \sysname.
The experiment results are visualized in the left two bars (\ie blue and orange bars) in each bar group in Figure~\ref{fig:operation-latency}.
For the AWS IoT, we use the cloud backend provided by the official AWS IoT Core services~\cite{awsiotcore} and an RPI as the local device. Note that the location of the cloud backend is automatically selected by the platform.
For the \sysname, we use the Intel Core~i7 laptop as the local controller and the same RPI as the device.
Both experiments use the same home WiFi network.

As shown, there is no significant difference between the bootstrapping operation, but the advantage of keeping communication local in \sysname becomes clear when focusing on content and command delivery, where \sysname is about 62\% faster in content delivery and 42\% faster in command delivery than the AWS IoT.

\begin{figure}[t]
	\centering
	\includegraphics[width=0.35\textwidth]{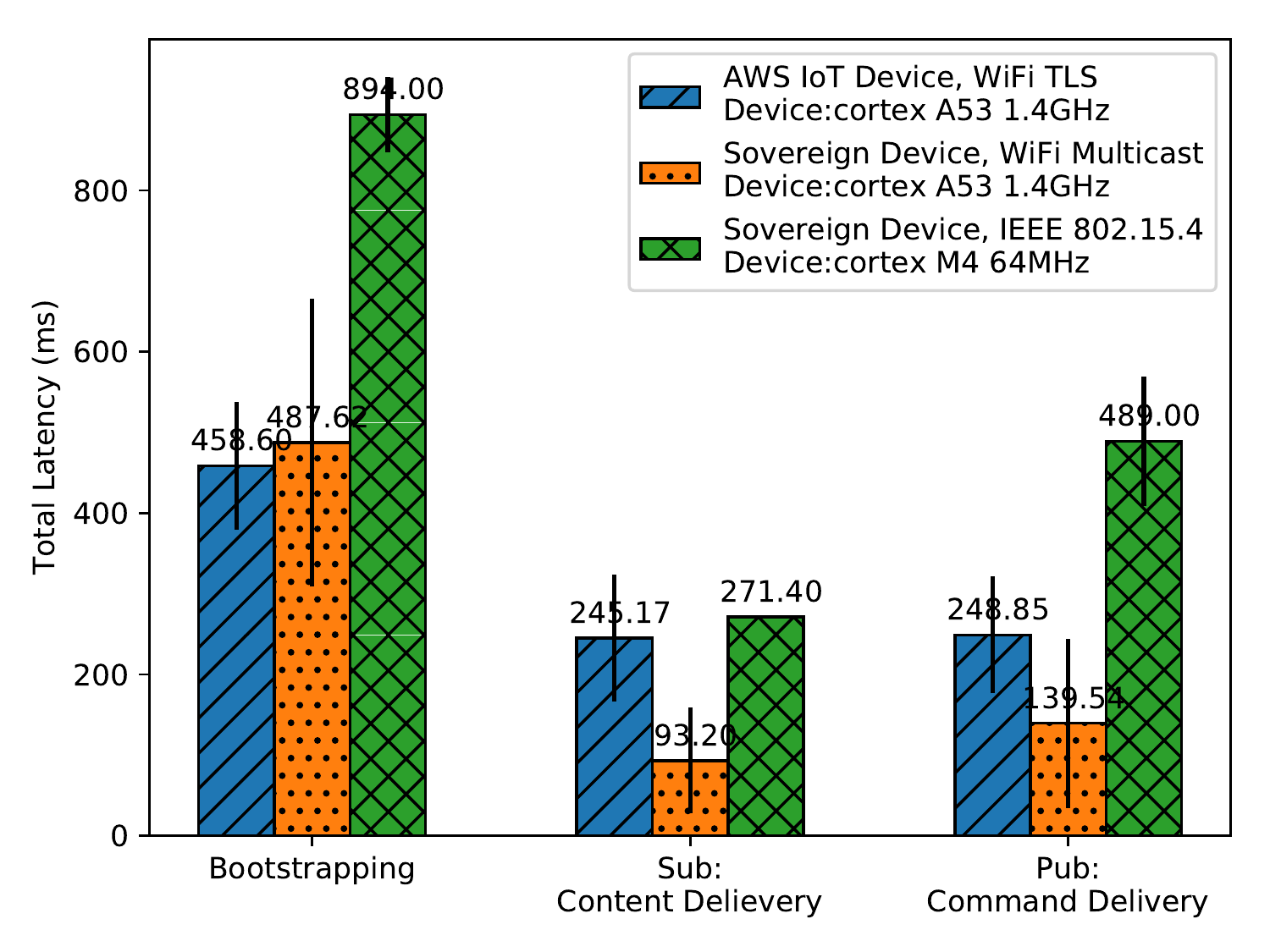}
	\caption{Latency in \sysname and Amazon AWS IoT}
	\label{fig:operation-latency}
\end{figure}

\mypara{Latency on Constrained Devices}
In addition, to measure \sysname's latency on constrained devices, we use the nRF52840 as the device and IEEE 802.15.4 network in our experiments.
The result is shown in the right bars (\ie green bars) of Figure~\ref{fig:operation-latency}.
As shown, it requires about 2-3x latency compared with the RPI implementation of \sysname.
However, since the bootstrapping operation is infrequent (\eg once in each device's lifetime) and the latency of content and command delivery are below 500~ms, we consider \sysname is still practical for the use of constrained devices in smart homes.

%

\mypara{Execution Time Breakdown}
A breakdown of \sysname's runtime into individual operations is provided in Figure~\ref{fig:micro-bench}. The visualized operations are performed when preparing Data before broadcasting to the network and Data processing after receiving. This includes digital signature creation and verification (ECDSA), content encryption and decryption (AES CBC), security policy checking, NDN packet encoding/decoding, and other cryptographic  operations (including ECDH, KDF). The results show that asymmetric cryptography consumes most of the computation time. This trend is observable across all evaluated devices.

\begin{figure}[t]
	\centering
	\includegraphics[width=0.48\textwidth]{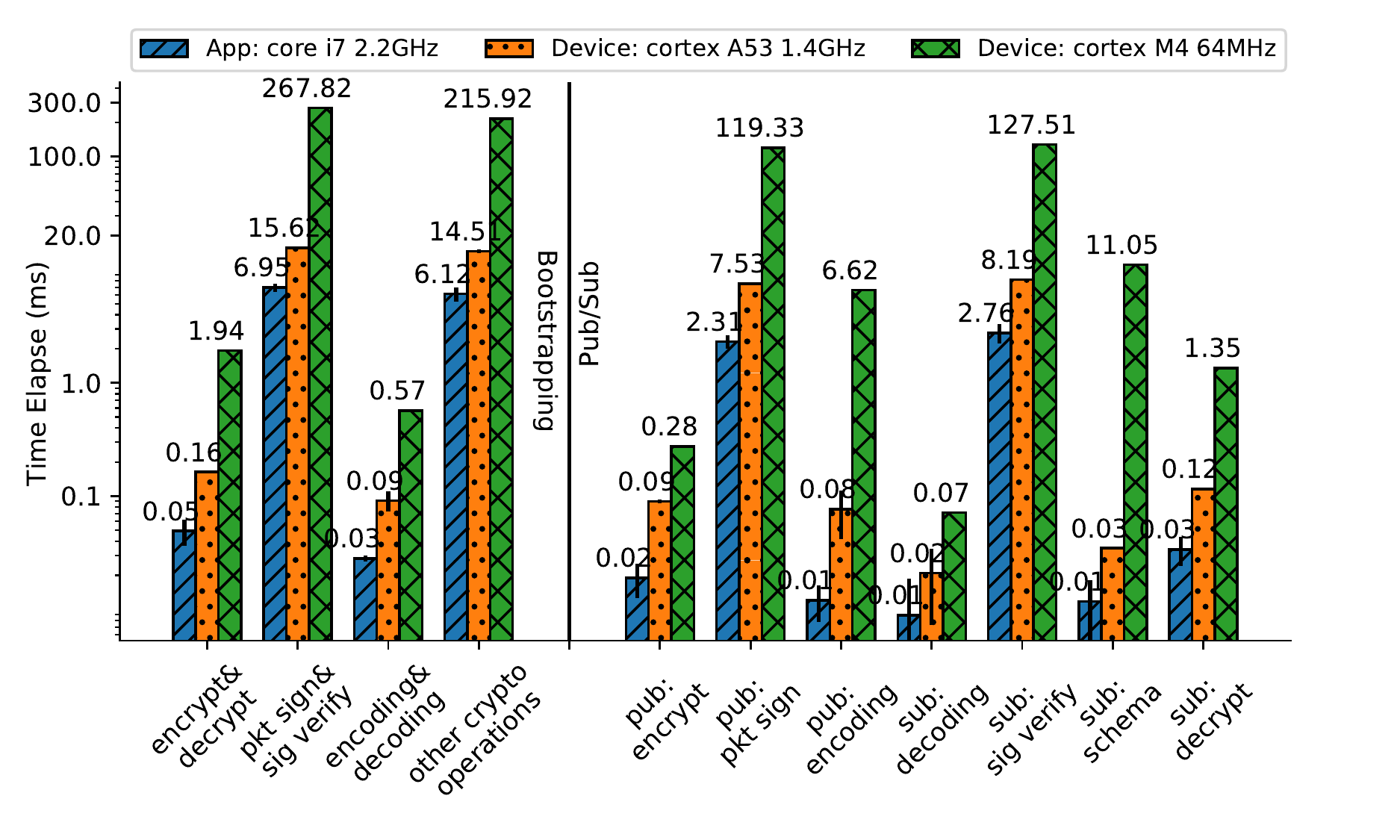}
	\caption{Breakdown of the Execution Time}
	\label{fig:micro-bench}
\end{figure}


\mypara{ROM and RAM Footprint}
We programmed an nRF52840 chip using RIOT OS~\cite{baccelli2013riot} for measuring \sysname's memory footprint. Table~\ref{table:evaluation-size} reports the size taken by individual \sysname modules. All the main modules together require less than 50~KB of RAM and 70~KB of ROM. This indicates \sysname's ability to be used on resource-constrained smart home devices.


\begin{table}[h]
	\footnotesize
	\resizebox{0.48\textwidth}{!}{
	\begin{tabular}{>{\centering\arraybackslash} m{2.5cm} || >{\centering\arraybackslash} m{2.5cm} | >{\centering\arraybackslash} m{2.5cm}}
		\hline
		Program/Modules & ROM Use &  RAM Use \\ \hline \hline
		\textbf{Subscriber in total} & \textbf{62KB}  &  \textbf{47.3KB} \\ \hline
		\textbf{Publisher in total} & \textbf{52.4KB} &  \textbf{38.2KB} \\ \hline \hline
		Application & 1.8\% & 7.3\% \\ \hline
		High-level Modules & 20.7\% & 34.2\% \\ \hline
		Utilities & 3.3\%  &  14.4\% \\ \hline
		Crypto Tools    & 25.1\% &  0.2\% \\ \hline
		Network Forwarder & 24.1\% &  25.0\% \\ \hline
		OS and Adaptation & 25.1\% &  18.9\% \\ \hline
	\end{tabular}
	}
	\smallskip
	\caption{ROM and RAM Consumption}
	\label{table:evaluation-size}

\end{table}

\subsection{Developer Friendliness Evaluation}
\label{subsec:eval_usability_user}

\begin{table}[t]
	\footnotesize
	\resizebox{0.48\textwidth}{!}{
	\begin{tabular}{>{\centering\arraybackslash} m{1cm}
							  || >{\centering\arraybackslash} m{4cm}
							  | >{\centering\arraybackslash} m{0.6cm}
							  | >{\centering\arraybackslash} m{0.6cm}
							  | >{\centering\arraybackslash} m{0.6cm}}
		\hline
		Stages & Descriptions & Avg. Time* & Min Time* & Max Time* \\ \hline \hline
		Tutorial  & Finish the \sysname tutorial & 68 & 20 & 90 \\ \hline
		Task~1 & Write an application that triggers an alarm when detecting smoke & 41 & 23 & 90 \\ \hline
		Task~2 & Write an application that controls an air conditioner based on temperature values & 24 & 15 & 40 \\ \hline
        Task~3 & Write an application that used motion sensor and lights to detect if someone is at home & 26 & 17 & 35 \\ \hline
	\end{tabular}
	}
	{\raggedright
	$*$: Counted in minutes \par
	}
%
	\vspace{5mm}
	\resizebox{0.48\textwidth}{!}{
	\begin{tabular}{>{\centering\arraybackslash} m{3cm}
						      || >{\centering\arraybackslash} m{1.3cm}
						      || >{\centering\arraybackslash} m{1.3cm}
						      || >{\centering\arraybackslash} m{1.3cm}}
		\hline
		Metrics & Avg. & Min & Max \\ \hline \hline
		Lines of Code & 72 & 15 & 114  \\ \hline
        Preparation Time** [min] & 39 & 10 & 90 \\ \hline
        Debugging Time [min] & 32 & 5 & 60 \\ \hline
	\end{tabular}
	}
	{\raggedright
		$**$: The time spent for reading documentation after finishing the tutorial. \par
	}
	\caption{Results of the Programming Experiment}
	\label{table:experience-summary}
\end{table}

%
%

As shown by the line-by-line comparison between the \sysname and SmartThings applications in Figure~\ref{fig:code-example}, it becomes apparent that \sysname offers a similar experience to developers as SmartThings.
Specifically, the highlighted lines reveal a high API similarity and indicate that developers can port existing smart home applications easily to \sysname without requiring changes to the application's logic.

We evaluated the developer friendliness by a programming experiment\footnote{Our study has been certified as an exempt study by UCLA Research Administration with protocol ID IRB\#20-001611.}.
We recruited computer science students as participants and asked each participant to write three new sample applications, including related device drivers, and to finish a questionnaire afterward.
We give no assistance except for providing a brief introduction, instructions, and the SDK documentation with code examples.
Appendix~\ref{appendix:survey} provides additional details on the programming experiment.

We found ten participants for the final programming experiment, none of which had prior experience in developing smart home software.
Among them, two participants experienced issues with the controller's graphical UI, and hence, did not finish the experiment.
Table~\ref{table:experience-summary} summarizes the results of the remaining eight participants.
As shown, the tutorial was finished by all participants in an average time of 68 minutes.
The average time for completing the first task was 41 minutes, and the remaining two tasks were solved in less time. We expect the speedup to be caused by a learning effect.

After completing the programming tasks, all participants finished an online questionnaire about \sysname's ease of use.
Two participants expressed concerns regarding the vague specification of function parameters and the long function names that are challenging to remember. 
Also, four participants suggested adding more illustrations about the system logic in the documentation.
Other than that, the participants faced no difficulties in using \sysname's SDK.

\section{Related Work}
\label{sec:relatedwork}

\begin{table}[t]

\newcolumntype{G}{>{\centering\let\newline\\\arraybackslash\hspace{0pt}}m{0.3\linewidth} }
\newcolumntype{H}{>{\centering\arraybackslash} m{0.25\linewidth} }

	\centering
	\footnotesize
	\begin{adjustbox}{max width=0.48\textwidth}
		\begin{tabular}{ G | H H H}
			\toprule
			\bf System & \bf Cloud Independent Home & \bf Fine-grained Security Policy  &  \bf Resilience to controller failure  \\
			\midrule
			HomeOS~\cite{homeOS} & Conditionally* & \yestick & \notick \\
			IoTGuard~\cite{celik2019iotguard} & N/A & \yestick & \notick \\
			SmartAuth~\cite{tian2017smartauth} & N/A & \yestick & \notick \\
			HanGuard~\cite{demetriou2017hanguard} & N/A & \notick & \notick \\
			ContextIoT~\cite{jia2017contexlot} & N/A & \yestick & \notick \\
			EXPAT~\cite{yahyazadeh2019expat} & N/A & \yestick & \notick \\
			IoT-IDM~\cite{nobakht2016host} & N/A & \notick & \notick \\
			SMP~\cite{sivaraman2015network} & N/A & \notick & \notick \\
			OpenHAB~\cite{openhab} & Conditionally** & \notick & \notick \\
			Home Assistant~\cite{home-assistant} & Conditionally** & \notick & \notick \\
			\midrule
			\rowcolor{verylightgray}
			\textbf{\sysname} & Fully & \yestick & \yestick  \\
			\bottomrule
	\end{tabular}
	\end{adjustbox}
	{
		\raggedright
		N/A$\colon$ The work is not relevant to whether the home relies on the cloud. \\
		*$\colon$ Cloud independent, when the controller is implemented locally. \\
		**$\colon$ Cloud independent, when purchased devices do not rely on the cloud. \\
		\par
	}
	\caption{Comparison of Representative Related Work.}
	\label{tab:related}
\end{table}

\mypara{Smart Home User Control}
Many works aim to enhance user control to today's smart home systems.
For example, HomeOS~\cite{homeOS} was proposed before the booming of the commercial smart home market.
Its main goal is to hide device heterogeneity from application development by treating all IoT devices as directly attached peripherals to a home PC.
However, the high dependency on a single centralized PC raised the concern of single point of failures; if one were to move that device to the cloud, it would resemble today's cloud-based solutions.
After cloud-based smart home deployment became popular, different solutions have been proposed to add runtime enforcement of security policies for better user control.
Some focus on modifying existing IoT application programs, or adding additional data flow verification onto the existing cloud backend~\cite{jia2017contexlot, tian2017smartauth, demetriou2017hanguard, celik2019iotguard, yahyazadeh2019expat}.
Other solutions add user control at the network layer~\cite{demetriou2017hanguard, simpson2017securing, nobakht2016host, sivaraman2015network}, \eg by assigning dedicated managers sitting at the network gateway to control whether to block individual network packets based on user-defined policies (\eg through Software Defined Network).
Nevertheless, these approaches do not move away from the cloud-based system model and thus do not fully address the issues as mentioned in \S\ref{sec:motivation}.
A comparison between these approaches and \sysname is shown in Table~\ref{tab:related}.


Home automation projects that provide user control with a local hub, like OpenHAB~\cite{openhab} and Home Assistant~\cite{home-assistant}, are seemingly similar to \sysname's local control.
The hub in these projects can directly control those devices that are cloud-independent, thus realizing local user control to some extent.
However, these systems have a different goal from ours: they focus more on providing a uniform control interface for users and facilitating the delivery of users' commands.
Thus, they do not provide a systematic solution to securing the communication between devices and to enforce inter-device access control.
In addition, since all user commands go through the local hub, the hub is a single point of failure.

\mypara{Use of Names}
Proposals that utilize semantically meaningful names closely relate to our approach.
For example, Intentional Naming System (INS)~\cite{ins} utilized semantic names inspired from earlier works~\cite{estrin1999next, jacobson1995kill, oki1993information} to build an application overlay for information and resource discovery.
Bolt~\cite{gupta2014bolt} also uses semantic names from the application layer by abstracting data as a stream of time-tag-value records, in which the semantic tags are used for data management.
Besides, SemIoTic~\cite{yus2019abstracting} utilizes a semantically meaningful meta-model based on SOSA/SSN ontology to describe smart spaces.
These works use names to facilitate service discovery and data management at the application layer, which indicates the undesired overhead like the IP-name mapping and the complexity to bridge the gap between the data-centric security and secured TCP/IP channels.
IoT wireless protocols, like Bluetooth Low Energy (BLE)~\cite{heydon2012bluetooth} and ZigBee\cite{zigbee}, also define service identifiers rather than using opaque network addresses for communication.
Nevertheless, the identifiers in BLE and Zigbee are solely for the purpose of communication.
In contrast, the names in \sysname are not only used for identifying resources but also used to execute security and define security policies.


\mypara{Pub/Sub}
Regarding the related works in pub/sub, conventional realizations, such as MQTT~\cite{mqtt}, rely on a central message broker that handles message filtering and forwarding.
The message broker serves as the rendezvous point between publishers and subscribers, interconnecting both ends at the application layer over a network of channel-based communication models.
\sysname's pub/sub implementation uses NDN's name semantic for defining message topics. This allows publishers and subscribers to rendezvous by namespace over
broadcast network media, removing the need for a central broker.
Our implementation also differs from existing NDN-based pub/sub designs~\cite{zhang2017scalable, shangpubsub2019, nichols2019lessons}, which are designed to run over multihop networks and utilize a synchronization protocol between publishers and subscribers.
\sysname leverages the local broadcast network setting to remove the need and overhead for a synchronization protocol.

\mypara{Related Works in ICN}
Recent years have also seen researches~\cite{shang2016named, ascigil2017keyword,  shang2017breaking, zhang2018ndnot, gundougran2018ndn, chakraborti2018using, future-device-ndn-2019, shangpubsub2019, gasw-triwt-20, lgsw-cdsfr-20} on the direction of ICN based smart homes.
For example, Shang~\etal~\cite{shang2016named, shang2017breaking} analyze the functions of cloud in today's cloud-based smart home systems and discuss how NDN can be potentially utilized to provide a replacement of cloud to operate the smart home in a local environment.
Ascigil~\etal~\cite{ascigil2017keyword} explore the different strategies of using NDN names for data fetching and computation power placing in edge IoT networks.
Other works~\cite{gasw-triwt-20, lgsw-cdsfr-20} also investigate how to combine the NDN/ICN with existing protocols like LowPAN or CoAP for IoT networking.
Discussing potential benefits and applying NDN/ICN to specific IoT protocols, existing works are in an early stage of the direction.
In this paper, we present a systematic NDN-based smart home framework with concrete implementation, which we believe is a big step forward in this direction and helps to examine the benefits of using NDN/ICN in IoT scenarios.

Comparing \sysname's access control with existing NDN based access control solutions like Name-based Access Control (NAC)~\cite{zhang2018nac}, the similarity is that they both utilize naming conventions for automated key delivery.
However, in order to work with constrained devices, \sysname's access control approach is more lightweight by directly delivering sealed symmetric decryption keys to authorized entities without computation-intensive asymmetric key encryption schemes.

%


\section{Discussion}
\label{sec:discussion}

After presenting the design and evaluation of \sysname in the earlier parts of the paper, this section critically discusses our design choices.

\mypara{\sysname Controller and System Resiliency}
The controller failure does not stop the system's normal operation.
Our experiments show that \sysname can continue operating during failures of the smart home controller.

However, since \sysname's security design relies on the controller acting as a trust anchor,
a compromised controller means hijacking the trust anchor, bringing all vulnerabilities similar to the compromised root certificate of a system.
Hence, we highly recommend hardening the access to the controller, and developing means for quick compromise detection.
One can add special protection for the controller, such as two-factor authentication, the use of hardware TPM's, and build anomaly detection on top of our framework.
We recognize that such measures might interfere with the smart home system's usability for end users; they raise end users' security awareness.

As part of our future work, we plan to relax the single point of trust, for example, with secret sharing schemes~\cite{10.1145/359168.359176} or threshold signatures~\cite{thresholdSecTR}.
In this way, the trust is jointly held by multiple parties, namely, multiple controllers can be installed (\eg a hub device in the home, and the homeowner's smartphone). Such schemes allow that all security-critical actions need to be verified by more than a single party, reducing the impact of a single compromised controller.

\mypara{Resiliency to Packet Losses and Link/Node Failures}
Two other commonly encountered failures in a networked system are network packet losses and link or node failures.
In \sysname, data consumers are responsible for reliable data fetching.
If an Interest packet times out without receiving the requested data, the consumer retransmits the Interest.
In addition, \sysname also utilizes redundancy mechanisms to improve packet delivery success:
\first fully utilizing broadcast media: if a device misses a notification message, it may hear a related response message issued by another device which received the notification;
and \second redundant notification: when sending notification for a newly available Data packet (\ie a command), the entity may send redundantly by retransmitting aggressively.
Regarding a failed or compromised devices in \sysname, the damage is limited.
A failed node does not affect others' direct communication, and even a compromised node cannot access resources that it is not allowed to.
When such a device is identified (\eg by applying existing intrusion detection mechanisms~\cite{celik2018soteria, celik2019iotguard, birnbackpeeves19, chi2020cross, fuhawather2021} above \sysname), its identity, as well as access rights, can be revoked by updating security policies and not renewing its keying material.



\mypara{Using the Cloud in \sysname}
\sysname's main contribution to user privacy protection is putting the \emph{control} of user data in the user's hand.
This does not exclude the use of cloud services.
A smart home system is free to outsource individual services such as backup storage, or computationally intensive tasks such as voice recognition, to cloud providers.
\sysname supports this by handling remote services as normal system entities. Same as local entities, cloud services need to obtain permission from the home controller to function or to access home data.



\mypara{Deployment of \sysname}
The deployment of smart home solutions is far more than a technical question and reaches beyond the scope of scientific research.
The goal of \sysname is to demonstrate an alternative to today's practice, and to encourage more discussion and development around this new direction.
It is noteworthy that \sysname does not depend on a global NDN deployment.
Even the deployment of NDN in the home network is transparent because the communication specifics are included in, and automatically handled by, the \sysname core library.
When NDN based designs become gradually adopted on edge systems and applications and bring desired benefits to end users, they can be expected to drive broader adoptions of NDN at a larger scale.

\section{Conclusion}
\label{sec:conclusion}

We make three observations based on our investigation into the \sysname design and development.
First, although cloud-based applications have achieved great successes, home IoT systems should take a different direction; putting the home system control in end users' hands offers a sure way to fully protect user privacy.
Second, the biggest challenge in designing a user-controlled home IoT system is security, and today's network security practices do not fit the home problem space.
Third, our experience with \sysname suggests that the new networking direction pointed out by ICN/NDN research, using semantically named and secured data as the basic system building block, shows great promise in enabling user-controlled smart home systems.

The \sysname design and development are still in an early stage with a few important pieces remain to be done.
In addition to the task of handling the controller failure, another urgent task is the design and integration of a local repository for logging data from smart home operations, which can be used for operation auditing and data analysis.  The reliability of such a local data storage can be further enhanced by saving an encrypted copy of the data in cloud-based storage systems as a backup.

We hope that this paper can serve as an invitation to all interested parties in joining us in exploring and experimenting with this new way of building smart homes.


%

\section*{Acknowledgment}
This work is partially supported by the National Science Foundation under award CNS-1629922 and CNS-1719403.

\appendices

\section{The Cloud in Existing Smart Home Systems}
\label{appendix:cloud}

In this appendix we use Amazon AWS IoT~\cite{awsiot}, Google IoT Core~\cite{google-assistant}, and Microsoft Azure IoT~\cite{microsoftiot} as examples to demonstrate the use of cloud backend in smart home systems.

%
%
%
%

\mypara{Amazon AWS IoT}
AWS IoT consists of three main components: cloud-connected devices, cloud-hosted applications, and the AWS Cloud.
In an AWS IoT home system, devices and applications must connect to the AWS Cloud through TLS with mutual authentication.
Therefore, each device and application must install two public key certificates in advance: one is AWS's certificate as the trust anchor and the other one is device/application certificate, which was issued when developers registered their products at AWS.
The cloud serves as the message broker and the authority to manage the system -- any unauthorized access to home resources will be rejected by the cloud.
Though the recent AWS Greengrass framework~\cite{awsgreengrass} encourages local communication, the management is still realized at the cloud.

%

%

\mypara{Google Assistant and Google IoT Core}
Google Assistant and Google IoT Core in general manage home/IoT system with Google's cloud services and OAuth.
Specifically, the cloud creates a database for each home containing the information of the structure, rooms, and devices.
This database serves as a global view of the home and is queried whenever there is an intent (\eg turn on the light).
For a device or an application to access some services, they need to obtain OAuth tokens from a OAuth server running in the cloud.
Besides, devices are implemented with code that is deployed as a webhook in the cloud.
When users send a command to the device, the command is first processed by the webhook and then forwarded back to the device at home.

\mypara{Microsoft Azure IoT}
Azure IoT uses the cloud to interact with individual devices. When receiving data, analysis will be performed at cloud side which is connected to other Azure cloud services. In D2D scenario, cloud will also act as a message broker between devices.
In the middle of cloud backend and devices, a predefined cloud gateway called Azure IoT Hub is involved. Azure IoT Hub has the capabaility of identity management for devices.
When connecting to the cloud, the device and cloud will be mutual authenticated by a TLS-based handshake.

\section{Developer Friendliness Study}
\label{appendix:survey}

The programming experiment discussed in Section~\ref{subsec:eval_usability_user} is designed as follows.
Participants were recruited via email and also received links to the briefing material in the same email.
The material includes brief background information on the project, experiment instructions, the SDK documentation, two device driver templates, and a sample tutorial application.
Beyond that, participants received no further assistance.
We asked each participant to go through our tutorial, try provided code templates, write three new sample applications with related device drivers, and complete a questionnaire during the experiment.
We set a time limit by telling the participants to exit the experiment when having spent more than 120 minutes.
The three sample applications aim to implement functionality as found in typical smart homes.
The first application raises the alarm based on the smoke detector's report.
The second monitors the ambient temperature and turns on the air-conditioner when the received temperature is above 80 degrees.
The third application turns off all lights in the living room when the motion sensor does not detect any motion for more than five minutes.

For evaluating the experiment design, we recruited two volunteers for pilot testing.
Minor improvements were made after evaluating the pilot tests, mainly in regards to the API documentation and clarity of the experiment instructions.
We want to note that participants took part in the experiment remotely and, therefore, we had no control over the participant's work environment.
However, we assume computer science students having an appropriate office workspace and, hence, only a minor impact on the experiment results.

The questionnaire contains four parts.
Each of them was asked at different stages of the experiment.
Part 1 surveys background information and was asked before starting the experiment.
Part 2 is asked between the tutorial and the actual programming.
Part 3 is asked after the programming experiment. The last part focuses on difficulties when using the API, asked after the experiment. Table~\ref{tab:questionnaire} lists all questionnaire questions.



\newcounter{qnum}\newcounter{snum}
\setcounter{qnum}{0}\setcounter{snum}{0}

\begin{table} [h]
\centering

\newcolumntype{H}{>{\arraybackslash} m{1cm} }
\newcolumntype{G}{>{\arraybackslash} m{7cm} }

\begin{tabular}{ c || G }

\bf No. & \bf Questions \\
\hline\hline

Part 1 & \emph{Background} \\
\hline\hline
1.1 & Do you have experience on home automation? \\
\hline
1.2 & If you have experience on home automation, what's the advantages/disadvantages of \sysname comparing to the home automation library you used before \\
\hline\hline

Part 2 & \emph{Learning} \\
\hline\hline
2.1 & How many minutes did it take for you to finish the quickstart example/tutorial? \\
\hline
2.2 & How many minutes did it take to learn the library before you begin coding? \\
\hline\hline

Part 3 & \emph{Programming Experiment} \\
\hline\hline
3.1 & How many minutes did it take for you to finish your task 1? \\
\hline
3.2 & How many minutes did it take for you to finish your task 2? \\
\hline
3.3 & How many minutes did it take for you to finish your task 3? \\
\hline
3.4 & How many minutes did it take to debug your application? \\
\hline
3.5 & How many lines of code did you write for your application? \\
\hline\hline

Part 4 & \emph{Summary and Suggestion}  \\
\hline\hline
4.1 & Do you face any difficulty when using the API? \\
    \hline
4.2 & Do you have any suggestion to improve the API? \\
\end{tabular}

\caption{Questionnaire accompanying the programming experiment}
\label{tab:questionnaire}
\end{table}



\bibliographystyle{IEEEtran}
\bibliography{reference}
%

%
%
%
%
%
%




\end{document}